\documentclass[letterpaper,twocolumn,10pt]{article}
\usepackage{hegroup}
\usepackage{graphicx}  
\usepackage{booktabs}  
\usepackage{multirow} 
\usepackage{colortbl}  
\usepackage[english,bidi=default]{babel}
\babelfont{rm}{TeXGyreTermesX} 
\babelprovide[import]{hindi}
\babelfont[*devanagari]{rm}{Lohit Devanagari}
\babelprovide[import]{arabic}
\babelfont[*arabic]{rm}{Noto Sans Arabic}
\usepackage{amsmath}
\usepackage{amssymb}
\usepackage{mathtools}
\usepackage{graphicx}
\usepackage{subcaption}
\usepackage{booktabs}
\usepackage{multirow}
\usepackage{makecell}
\usepackage{array}
\usepackage{xcolor}
\usepackage{xspace}
\usepackage{enumitem}
\usepackage{natbib} 
\usepackage{algorithm}
\usepackage{algorithmic}
\usepackage[capitalize,noabbrev]{cleveref}

\newcommand{\mypara}[1]{\smallskip\noindent\textbf{#1.}\xspace}

\title{Beyond Waveform Robustness: Robust Feature-Vocoder Adversarial Attacks on Automatic Speech Recognition }

\date{}

\author{
Yifan Liao\textsuperscript{1,2} \ \
Zongmin Zhang\textsuperscript{1} \ \ \
Zhen Sun\textsuperscript{1} \ \ \
Yuhui Sun\textsuperscript{1} \ \ \
Xinhu Zheng\textsuperscript{1} \ \ \
Xinlei He\textsuperscript{2}\thanks{Corresponding author: Xinlei He (\href{mailto:xinlei.he@whu.edu.cn}{xinlei.he@whu.edu.cn})} \ \ \
\\
\\
\textsuperscript{1}\textit{The Hong Kong University of Science and Technology (Guangzhou)} \ \ \\
\textsuperscript{2}\textit{Wuhan University} \ \ \\
}

\begin{document}

\maketitle

\begin{abstract}
Automatic speech recognition (ASR) systems have become widely used for multilingual speech-to-text transcription. 
Their robustness to adversarial attacks has become an important topic for the community.
Existing adversarial attacks directly add adversarial noise to the speech audio.
However, prior work has shown that existing adversarial attacks face two limitations: they often transfer poorly to black-box ASR systems and are increasingly mitigated by defenses tailored to input-space perturbations.
In this work, we propose a \textbf{Clean-Referenced Feature-Vocoder Attack}, a surrogate-based black-box attack that moves the adversarial search space from raw waveforms to self-supervised learning (SSL) representations. 
To address the transferability limitation, we perturb more generalizable acoustic-phonetic representations rather than low-level waveform samples, reducing dependence on surrogate-specific waveform gradients and encouraging adversarial perturbations that generalize across ASR systems.
To bypass different defenses, we shift the adversarial signal from explicit additive waveform noise to SSL feature-space perturbations and reconstruct them through a vocoder into speech-like waveform adversarial signals, making the resulting samples less aligned with waveform-bounded defenses.
Extensive experiments show that, when optimized only on raw Whisper-small as a public surrogate model, our attack transfers effectively to black-box ASR models with a +26.6 WER improvement over the SOTA baseline, while also remaining effective against multiple training defenses with a +36.2 WER improvement.
These results reveal a blind spot in current ASR robustness evaluation. 

\end{abstract}

\section{Introduction}
\label{sec:introduction}

\begin{figure*}[t]
    \centering
    \includegraphics[width=0.9\linewidth]{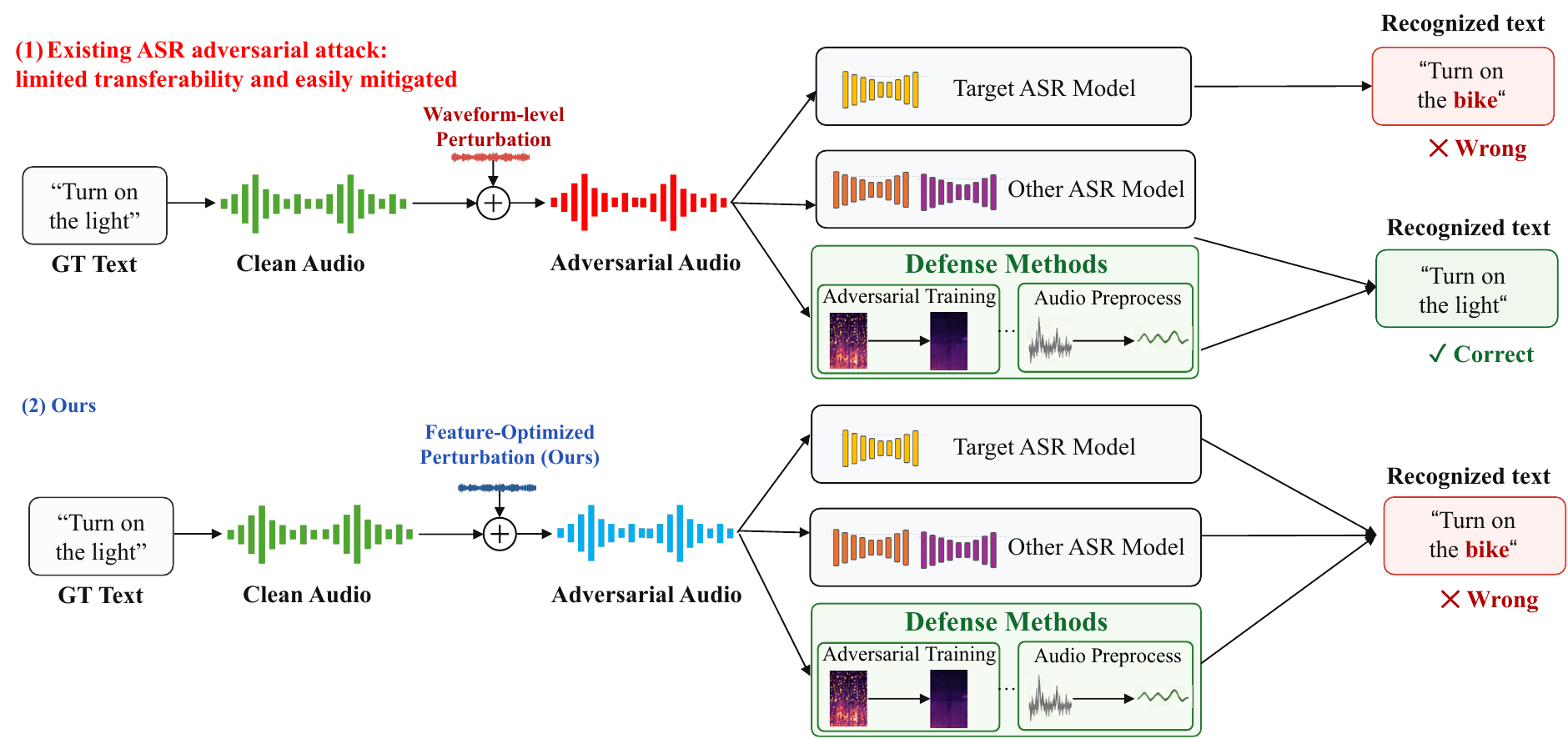}
    \caption{
    \small Traditional waveform-level attacks directly add perturbations to the input audio, which can induce transcription errors on target ASR models but often suffer from limited transferability and are increasingly addressed by defenses such as adversarial training. In contrast, our method optimizes adversarial perturbation in the feature space, improving transferability across ASR models and maintaining effectiveness against defended ASR systems.
    }
    \label{fig:asr_intro}
\end{figure*}

Automatic speech recognition (ASR) has become a mainstream technology for converting spoken audio into text~\citep{DBLP:journals/corr/HannunCCCDEPSSCN14, DBLP:conf/nips/BaevskiZMA20}. 
Recent speech foundation models, such as Whisper improve the generality of ASR systems by
supporting multilingual speech-to-text transcription across diverse acoustic conditions~\citep{radford2023whisper,baevski2020wav2vec}.
As ASR systems are increasingly deployed in real-world applications, their robustness to adversarial attacks has become an important concern for the communities.

The goal of adversarial attack on ASR is to induce transcription errors while preserving the speech quality of the original audio.
~\Cref{fig:asr_intro} illustrates the scenario: adversarial attack on ASR can cause incorrect recognition.
Existing attacks typically achieve this by adding small perturbations to the input waveform~\citep{DBLP:conf/sp/Carlini018, DBLP:conf/icml/QinCCGR19, DBLP:conf/interspeech/NeekharaHPDMK19,raina2024muting,raina2024controlling}. 
While these attacks reveal important vulnerabilities on ASR, according to prior work in other domains~\cite{zhang2024constructing,li2024transcending}, these attacks constrained to the input space often suffer two inherent limitations.
First, they often suffer from limited black-box transferability because they optimize low-level, sample-wise perturbations in the raw audio space.
These perturbations can overfit to surrogate-specific gradients and may not generalize to unseen ASR systems.
Second, waveform-level perturbations are increasingly addressed by existing defenses because the adversarial information is explicitly represented as additive waveform noise.
This makes the attack align with the assumptions of waveform-oriented defenses, such as input preprocessing and adversarial training defenses against waveform-bounded perturbations.~\cite{dong2019evading,hussain2021waveguard,olivier2021sequential,wu2023defending}

To address these two limitations, we propose \textbf{Clean-Referenced Feature-Vocoder Attack}, a surrogate-based black-box attack against defended ASR models. 
Firstly, to address the transferability limitation, we avoid optimizing low-level sample-wise perturbations on the raw waveform. 
Instead, we perturb intermediate self-supervised learning  (SSL) representations~\cite{chen2022wavlm}, which encode higher-level acoustic and phonetic information shared across ASR systems. 
This reduces the dependence on surrogate-specific waveform gradients and encourages adversarial perturbations that can generalize across different ASR models. 
To bypass current defense methods, we reconstruct the perturbed representation back into audio using a frozen neural vocoder~\cite{kong2020hifi}, rather than adding the perturbation directly to the waveform. 
In this way, the adversarial perturbation is embedded through a feature-to-waveform speech reconstruction process, making it less aligned with waveform-bounded defenses such as input transformations and adversarial training.

Extensive experiments show that our attack reveals a blind spot in current ASR robustness evaluation. 
Optimized only on raw Whisper-small as a public surrogate model, our attack transfers effectively to black-box ASR models across both Whisper-family variants and CTC-based architectures, outperforming the strongest baseline by an average of +26.6 WER.
It also remains effective against multiple adversarial-training defenses, achieving an average improvement of +36.2 WER over the SOTA baseline on English dataset and +31.3 CER over the SOTA baseline on Chinese dataset.
Moreover, our attack maintains strong performance under input preprocessing defenses, showing that feature-vocoder adversarial examples are not well covered by defenses designed for waveform-bounded perturbations. 
These results indicate that robustness against additive waveform noise or adversarial prefixes can overestimate the security of ASR systems.

Our contributions are as follows:
\begin{itemize}
    \item We introduce a clean-referenced feature-vocoder attack space for ASR, where adversarial examples are generated by perturbing SSL speech features rather than by adding waveform noise.
    \item We design a clean-referenced feature-vocoder framework that embeds adversarial variation into SSL speech representations and reconstructs them into natural waveform audio through a frozen vocoder instead of adding explicit waveform noise.
    \item Extensive experiments show that adversarial audio optimized on raw Whisper-small transfers strongly to multiple defense methods, revealing a blind spot of waveform-oriented ASR robustness.
\end{itemize}

\section{Related Work}
In recent years, large-scale weakly supervised ASR foundation models, exemplified by Whisper~\cite{radford2023whisper}, have substantially improved the generalization ability and practical robustness of speech recognition across multilingual, multitask, and naturally noisy settings~\cite{radford2023robust}.
However, prior studies have shown that robustness under natural conditions does not automatically translate into adversarial robustness.
Early work demonstrated that end-to-end ASR systems can be steered toward target transcriptions by carefully crafted small waveform perturbations with little impact on human perception~\cite{carlini2018audio}; subsequent studies further advanced such attacks toward settings that simultaneously emphasize psychoacoustic imperceptibility and environmental robustness~\cite{qin2019imperceptible}.
In the foundation ASR setting, Whisper has likewise been shown to remain highly vulnerable to adversarial examples~\cite{olivier2022there}, and recent attacks have evolved from per-example optimization to more practical universal acoustic prefixes and controllable triggers that can induce muting or manipulate model outputs~\cite{raina2024muting,raina2024controlling}.
Besides, recent work has shown that the adversarial attack targeting the SSL model can be robust~\cite{liao2026escaping}.
Correspondingly, existing defenses have mainly focused on input preprocessing based on audio transformations and consistency checking, sequential randomized smoothing, and generative purification~\cite{hussain2021waveguard,olivier2021sequential,wu2023audiopure}, but these methods are still largely developed under a waveform-level perturbation threat model.
On the other hand, modern voice conversion research suggests that high-quality speech generation increasingly relies on a representational manifold jointly constrained by SSL representations and neural vocoders: large-scale speech representations such as WavLM provide a unified foundation for modeling content and speaker-related attributes~\cite{chen2022wavlm}, kNN-VC shows that high-quality any-to-any conversion can be achieved through local neighborhood replacement in the representation space alone~\cite{baas2023voice}, and ACE-VC further demonstrates the stronger controllability enabled by explicit disentanglement~\cite{hussain2023ace}; meanwhile, HiFi-GAN and subsequent timbre-aware vocoders make it possible to stably map intermediate representations back to natural waveforms~\cite{kong2020hifi,guo2024vec2wav}.
Building on these advances, we argue that voice conversion should not be viewed merely as a speech editing tool, but also as a natural speech manifold jointly shaped by representation-space structure and the reconstructor.
This provides a direct motivation for shifting adversarial example search from arbitrary waveform noise to content-preserving generation within a VC manifold, and it also serves as the starting point of this work for re-examining the robustness boundary of adversarially trained ASR.

\section{Threat model}
We study a surrogate-based black-box attack against ASR systems. Let $f_t$ denote the target ASR model and $f_s$ denote a public surrogate model. 
The adversary has white-box access to $f_s$  but has no access to $f_t$. 
During attack generation, the target model $f_t$ is not used.

Given a clean speech audio $x$ with ground-truth transcript $y$, an ASR model maps the input waveform to a predicted token sequence:
\begin{equation}
    \hat{y}_s = f_s(x), 
    \qquad 
    \hat{y}_t = f_t(x).
\end{equation}
We evaluate transcription failure using an error-rate function $\operatorname{Err}(\hat{y}, y)$, such as WER for English or CER for Chinese.

The goal of the attacker is to construct an adversarial example  $x_{\mathrm{adv}}$ that transfers from the surrogate model to the target model, causing a high error rate on $f_t$ while preserving the perceived speech content and quality of the original audio. Since $f_t$ is black-box, this goal is implemented by optimizing only a surrogate loss on $f_s$.

In our feature-vocoder attack, the feasible set is defined by perturbing the SSL feature representation of the clean audio and reconstructing the perturbed feature through a frozen vocoder. Let $E$ be the frozen SSL encoder and $V$ be the frozen vocoder. We generate
\begin{equation}
    x_{\mathrm{adv}}(\delta) = V(E(x) + \delta),
\end{equation}
where $\delta$ denotes the feature-space perturbation

After generation, transferability is evaluated by measuring
\begin{equation}
    \operatorname{Err}\bigl(f_t(x_{\mathrm{adv}}), y\bigr)
\end{equation}
on the black-box target models and defenses.

\begin{figure}[t]
    \centering
    \includegraphics[width=0.95\linewidth]{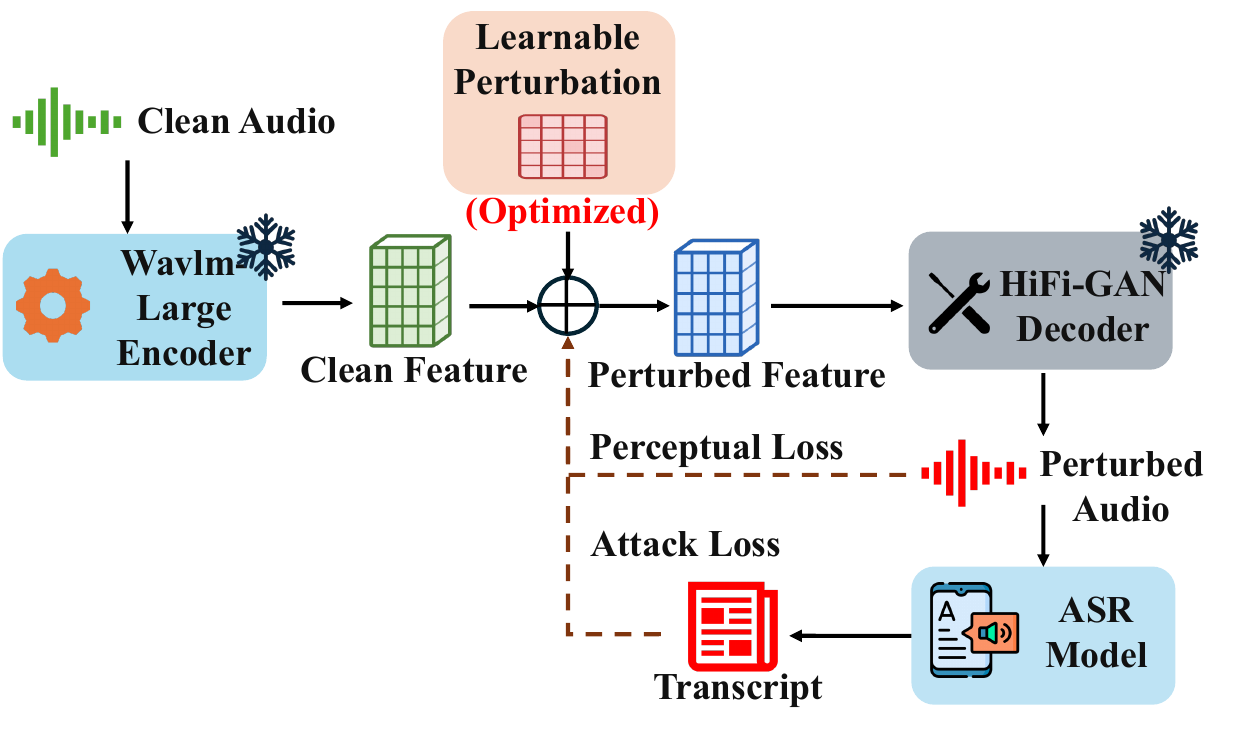}
    \caption{
    Overview of the proposed Clean-Referenced Feature-Vocoder Attack. 
    }
    \label{fig:asr_adv_method}
\end{figure}

\section{Method}
\label{sec:method}

We propose a clean-referenced feature-vocoder attack for evaluating the robustness of ASR systems beyond waveform-bounded perturbations.
The pipeline is shown in~\Cref{fig:asr_adv_method}.
Given a clean speech audio, a frozen encoder extracts the SSL feature. 
We optimize a learnable perturbation in the SSL feature space and reconstruct the perturbed feature trajectory into waveform audio using a frozen vocoder. 
The generated audio is fed into the surrogate ASR model, and the ASR loss together with a perceptual regularizer loss are back-propagated to update the feature-space perturbation.

\paragraph{Feature extraction.}
Given a clean speech audio $x$ with ground-truth transcript $y$, we use a frozen SSL speech encoder $E(\cdot)$ to extract a frame-level representation:
\begin{equation}
    q = E(x), \qquad q = (q_1,\ldots,q_T).
\end{equation}
Here, $q_t$ denotes the SSL feature at frame $t$. The encoder is frozen during attack optimization.

\paragraph{Direct feature-space perturbation.}
Let $q=E(x)\in\mathbb{R}^{T\times D}$ denote the SSL feature trajectory of the clean audio, where $T$ is the number of frames and $D$ is the feature
dimension. We introduce a learnable feature-space perturbation $\delta\in\mathbb{R}^{T\times D}$ and define the perturbed feature trajectory as
\begin{equation}
    z(\delta) = q + \delta .
\end{equation}
To make the attack budget explicit, we constrain $\delta$ by a normalized
feature-space radius:
\begin{equation}
    \mathcal{U}_{\rho}(x)
    =
    \left\{
    \delta:
    \frac{\|\delta\|_F}{\|q\|_F+\epsilon}
    \leq \rho
    \right\},
\end{equation}
where $\rho$ controls the perturbation scale and $\epsilon$ is a small constant for numerical stability. The adversarial waveform is reconstructed by a frozen vocoder:
\begin{equation}
    x_{\mathrm{adv}}(\delta)
    =
    V(z(\delta))
    =
    V(E(x)+\delta).
\end{equation}
Thus, the feasible attack space is
\begin{equation}
    \mathcal{M}_{\mathrm{FV}}(x;\rho)
    =
    \left\{
    V(E(x)+\delta):
    \delta \in \mathcal{U}_{\rho}(x)
    \right\}.
\end{equation}
Unlike waveform-level attacks, this constraint is imposed in the SSL representation space.

\paragraph{Attack loss.}
Let $f_s$ denote the surrogate ASR model used for optimization. We optimize the feature-space perturbation $\delta$ to reduce the surrogate model's probability of the correct transcript. Formally, we maximize the negative log-likelihood of the ground-truth transcript under the adversarial audio:
\begin{equation}
    \max_{\delta\in\mathcal{U}_{\rho}(x)}
    \operatorname{NLL}_{\mathrm{text}}
    \left(
    y \mid x_{\mathrm{adv}}(\delta); f_s
    \right).
\end{equation}
Equivalently, we minimize
\begin{equation}
    \mathcal{L}_{\mathrm{attack}}(\delta)
    =
    -
    \operatorname{NLL}_{\mathrm{text}}
    \left(
    y \mid x_{\mathrm{adv}}(\delta); f_s
    \right).
\end{equation}
We compute $\operatorname{NLL}_{\mathrm{text}}$ over lexical transcript tokens and exclude Whisper prompt, language, task, and other special tokens.

\paragraph{Clean-referenced perceptual loss.}
Directly optimizing feature-space perturbations may produce temporally unstable feature trajectories or abnormally high-frequency artifacts after vocoding.
We therefore use a clean-referenced perceptual regularizer. The perceptual loss is
\begin{equation}
    \mathcal{L}_{\mathrm{perc}}(\delta)
    =
    \frac{
        \operatorname{TV}(z(\delta))
    }{
        \operatorname{sg}(\operatorname{TV}(q))+\epsilon
    }
    +
    \alpha
    \frac{
        \operatorname{HF}(x_{\mathrm{adv}}(\delta))
    }{
        \operatorname{sg}(\operatorname{HF}(x))+\epsilon
    },
\end{equation}
where $\operatorname{sg}(\cdot)$ denotes stop-gradient normalization.

The first term penalizes temporal jitter in the perturbed SSL feature trajectory:
\begin{equation}
    \operatorname{TV}(z)
    =
    \frac{1}{T-1}
    \sum_{t=2}^{T}
    \|z_t-z_{t-1}\|_2^2.
\end{equation}
The second term penalizes the high-frequency energy in the generated audio:
\begin{equation}
    \operatorname{HF}(x)
    =
    \frac{
        \sum_{f>f_c}|S_f(x)|^2
    }{
        \sum_f |S_f(x)|^2
    },
\end{equation}
where $S_f(x)$ denotes the short-time Fourier transform coefficient at frequency bin $f$. The high-frequency term is normalized by the clean audio rather than by a reconstructed baseline, so that the regularizer directly references the frequency statistics of the original audio.

\paragraph{Final objective.}
The final optimization objective contains the surrogate attack loss and the clean-referenced perceptual regularizer:
\begin{equation}
    \delta^\star
    =
    \arg\min_{\delta\in\mathcal{U}_{\rho}(x)}
    \left[
    \mathcal{L}_{\mathrm{attack}}(\delta)
    +
    \lambda_{\mathrm{perc}}
    \mathcal{L}_{\mathrm{perc}}(\delta)
    \right].
\end{equation}
The final adversarial audio is then
\begin{equation}
    x_{\mathrm{adv}}
    =
    x_{\mathrm{adv}}(\delta^\star)
    =
    V(E(x)+\delta^\star).
\end{equation}
The attack loss drives the adversarial audio away from the correct transcript on the surrogate ASR model, while the perceptual regularizer discourages temporal instability and abnormal high-frequency artifacts. Gradients are back-propagated through the surrogate ASR model, the frozen vocoder, and the perturbed SSL feature trajectory to update only $\delta$.

\section{Experiments}

\mypara{Dataset and Models}
We evaluate the proposed method on two representative ASR datasets covering two languages. The first is LibriSpeech train-clean 100~\cite{DBLP:conf/icassp/PanayotovCPK15}, which contains approximately 100 hours of English read speech. 
The second is AISHELL-1~\cite{DBLP:conf/ococosda/BuDNWZ17}, a Mandarin Chinese ASR corpus with approximately 170 hours of speech. 

The target models are Whisper-family and SSL-based ASR models~\cite{radford2023whisper,guo2024vec2wav,hsu2021hubert}, and to assess the robustness of the attack under practical defensive settings, we further evaluate it against two categories of defenses. 
The first category consists of adversarial-training-based defenses, including Cross-Entropy adversarial training (CE-AT), Decoding Trajectory adversarial training (DE-AT), Dynamic Margin Weighting (DMW), and Translation-Invariant (TI)~\cite{madry2018towards,dong2019evading,liu2021probabilistic}. 
The second category consists of input-preprocessing-based defenses, including LPF, WaveGuard, AudioPure, and PVP Vote~\cite{kwon2020acoustic,hussain2021waveguard,wu2023audiopure,olivier2021sequential}.

\mypara{Baselines}
We compare our method with representative adversarial attacks, including PGD~\cite{madry2018pgd}, MI-FGSM~\cite{dong2018mifgsm}, VMI-FGSM~\cite{wang2021enhancing}, Muting Whisper~\cite{raina2024muting}, and SlothSpeech~\cite{haque2023slothspeech}. For fair comparison, all attacks are generated from the same clean utterances and optimized only on the same public surrogate model. We use raw Whisper-small as the surrogate for LibriSpeech and the AISHELL-1 fine-tuned Whisper-small surrogate for AISHELL-1.
All generated adversarial examples are then evaluated on the same black-box target models and defenses. The details can be reviewed in ~\ref{sec:appendix-implementation}

\begin{table*}[t]
\centering
\resizebox{\textwidth}{!}{
\begin{tabular}{l|c|cccc|cccc}
\toprule
\multirow{2}{*}{\textbf{Attack Method}} 
& \multirow{2}{*}{\textbf{Raw Model}} 
& \multicolumn{4}{c|}{\textbf{Adversarial Training Defenses}} 
& \multicolumn{4}{c}{\textbf{Input Preprocessing Defenses}} \\
\cmidrule(lr){3-6} \cmidrule(lr){7-10}
& & \textbf{CE-AT} & \textbf{DT-AT} & \textbf{DMW} & \textbf{TI} 
& \textbf{LPF} & \textbf{WaveGuard} & \textbf{AudioPure} & \textbf{PVP Vote} \\
\midrule
\multicolumn{10}{l}{\textbf{LibriSpeech (English)}} \\
\midrule
Clean (No Attack) 
& 4.75 & 5.72 & 5.70 & 5.86 & 5.81 
& 6.45 & 6.76 & 6.17 & 4.80 \\
PGD
& 61.06 & 8.69 & 8.13 & 9.61 & 9.80 
& 16.28 & 15.47 & 8.38 & 45.62 \\
MI-FGSM 
& 77.65 & 16.20 & 16.98 & 16.86 & 16.85  
& 39.24 & 32.25 & 27.33 & 76.11 \\
VMI-FGSM 
& 78.84 & 30.29 & 30.91 & 34.48 & 39.14 
& 42.04 & 55.36 & 42.38 & 78.34  \\
Muting Whisper
& \textbf{99.52} & 6.11 & 6.09 & 5.94 & 6.04  
& 7.52 & 7.62 & 6.80 & \textbf{79.52} \\
Sloth
& 39.50 & 9.38 & 8.71 & 9.26 & 9.38  
& 14.88 & 14.98 & 8.70 & 28.47 \\
\rowcolor{gray!10}
\textbf{Feature Attack (Ours)} 
& 75.43 & \textbf{71.26} & \textbf{67.23} & \textbf{70.61} & \textbf{70.63} 
& \textbf{68.39} & \textbf{57.03} & \textbf{70.86} & 78.60 \\
\midrule
\multicolumn{10}{l}{\textbf{AISHELL-1 (Chinese)}} \\
\midrule
Clean (No Attack) 
& 6.29 & 5.51 & 5.87 & 5.66 & 5.22  
& 10.03 & 7.74 & 27.67 & 5.75 \\
PGD
& 69.01 & 15.08 & 14.86 & 15.19 & 17.39  
& 26.30 & 12.67 & 26.71 & 64.27 \\
MI-FGSM 
& 79.87 & 23.29 & 23.72 & 28.33 & 23.25 
& 51.71 & 59.45 & 33.49 & 74.18 \\
VMI-FGSM 
& 80.36 & 31.26 & 31.90 & 35.58 & 40.39 
& 43.38 & 57.13 & 43.73 & \textbf{79.82} \\
Muting Whisper
& \textbf{94.47} & 5.93 & 6.05 & 5.87 & 5.41  
& 6.74 & 6.89 & 5.96 & 75.24 \\
Sloth
& 53.11 & 14.33 & 11.26 & 8.43 & 9.82  
& 36.04 & 31.92 & 20.24 & 54.29 \\
\rowcolor{gray!10}
\textbf{Feature Attack (Ours)} 
& 72.25 & \textbf{66.57} & \textbf{66.10} & \textbf{66.07} & \textbf{65.49} 
& \textbf{72.10} & \textbf{70.34} & \textbf{69.92} & 75.31 \\
\bottomrule
\end{tabular}
}
\caption{Performance of the proposed Feature Attack against various baseline attacks and defense mechanisms on the Whisper-small model. Performance is evaluated using WER (\%) for English and CER (\%) for Chinese.}
\label{tab:attack_results}
\end{table*}

\mypara{Metrics}
We use word error rate (WER, \%) for English speech recognition on LibriSpeech and character error rate (CER, \%) for Chinese speech recognition on AISHELL-1. Higher WER or CER indicates stronger attack effectiveness, as it reflects a larger degradation in the transcription quality of the ASR model. 

\mypara{Implementation Details}
We implement our attack using a frozen WavLM-Large model as the SSL feature encoder and a frozen HiFi-GAN vocoder for waveform reconstruction.  
For each clean audio, we extract its SSL feature trajectory and optimize a bounded feature-space perturbation $\delta$ for 50 steps. 
Unless otherwise specified, we set the normalized feature perturbation budget to $\rho=0.1$ and the clean-referenced perceptual regularization weight to $\lambda_{\mathrm{perc}}=1$. 
We set the $6000~\mathrm{Hz}$ for the $f_c$ in frequency energy loss.
For English experiments on LibriSpeech, we use raw Whisper-small as the surrogate model. 
For AISHELL-1, we initialize the surrogate model from Whisper-small and fine-tune it on the AISHELL-1 development set to obtain a Mandarin surrogate. 
The defended target models are trained on the AISHELL-1 training set, and all final attack results are reported on the held-out AISHELL-1 test set.
The test set is never used for surrogate fine-tuning and defense training.

\mypara{Effectiveness Results}
~\Cref{tab:attack_results} compares our Feature Attack with representative waveform-level attacks on Whisper-small under both adversarial training and input preprocessing defenses. Overall, existing attacks can be highly effective on the undefended raw model, but their performance drops sharply once defenses are applied. 
For example, Muting Whisper achieves the highest raw-model error rate, reaching 99.52\% WER on LibriSpeech and 94.47\% CER on AISHELL-1, but its error rate falls close to the clean baseline under adversarial training defenses. 
MI-FGSM is more robust than PGD and Sloth under some preprocessing defenses, but it still degrades substantially under adversarial training.

In contrast, our Feature Attack consistently maintains high error rates across defended models. 
For instance, on LibriSpeech, our method achieves 71.26\% WER under CE-AT and 70.86\% WER under AudioPure, while all baseline attacks are much less effective in these settings. 
On AISHELL-1, our attack similarly remains effective, achieving 66.57\% CER under CE-AT and 70.34\% CER under WaveGuard. 
This indicates that conventional waveform-level attacks are largely aligned with the assumptions of existing defenses, whereas our feature-vocoder attack exposes vulnerabilities that remain under both adversarial training and input preprocessing.

\begin{figure}[t]
    \centering
    \begin{subfigure}[t]{0.49\linewidth}
        \centering
        \includegraphics[width=\linewidth]{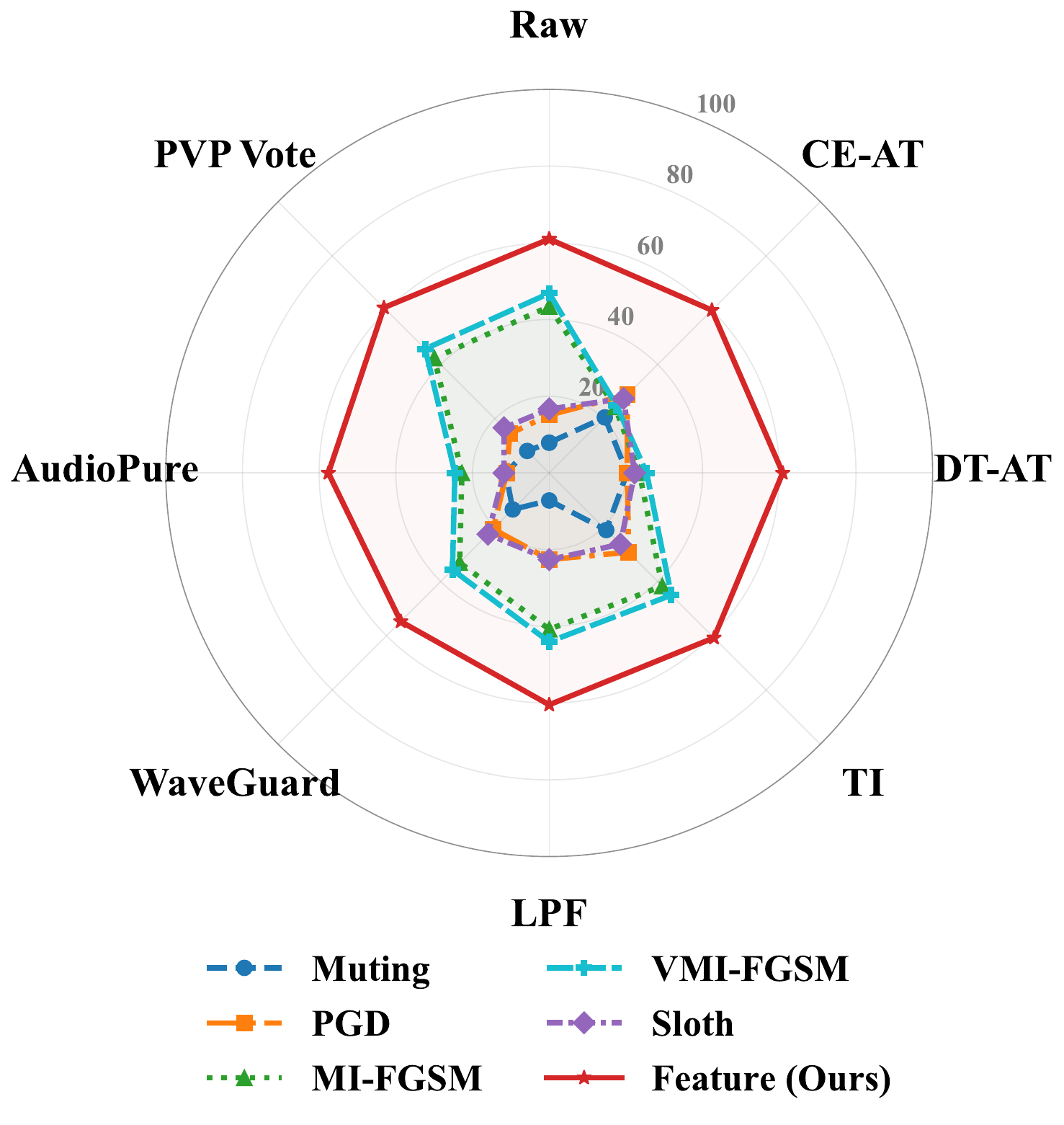}
        \caption{Whisper-tiny}
        \label{fig:whisper_transfer_tiny}
    \end{subfigure}
    \hfill
    \begin{subfigure}[t]{0.49\linewidth}
        \centering
        \includegraphics[width=\linewidth]{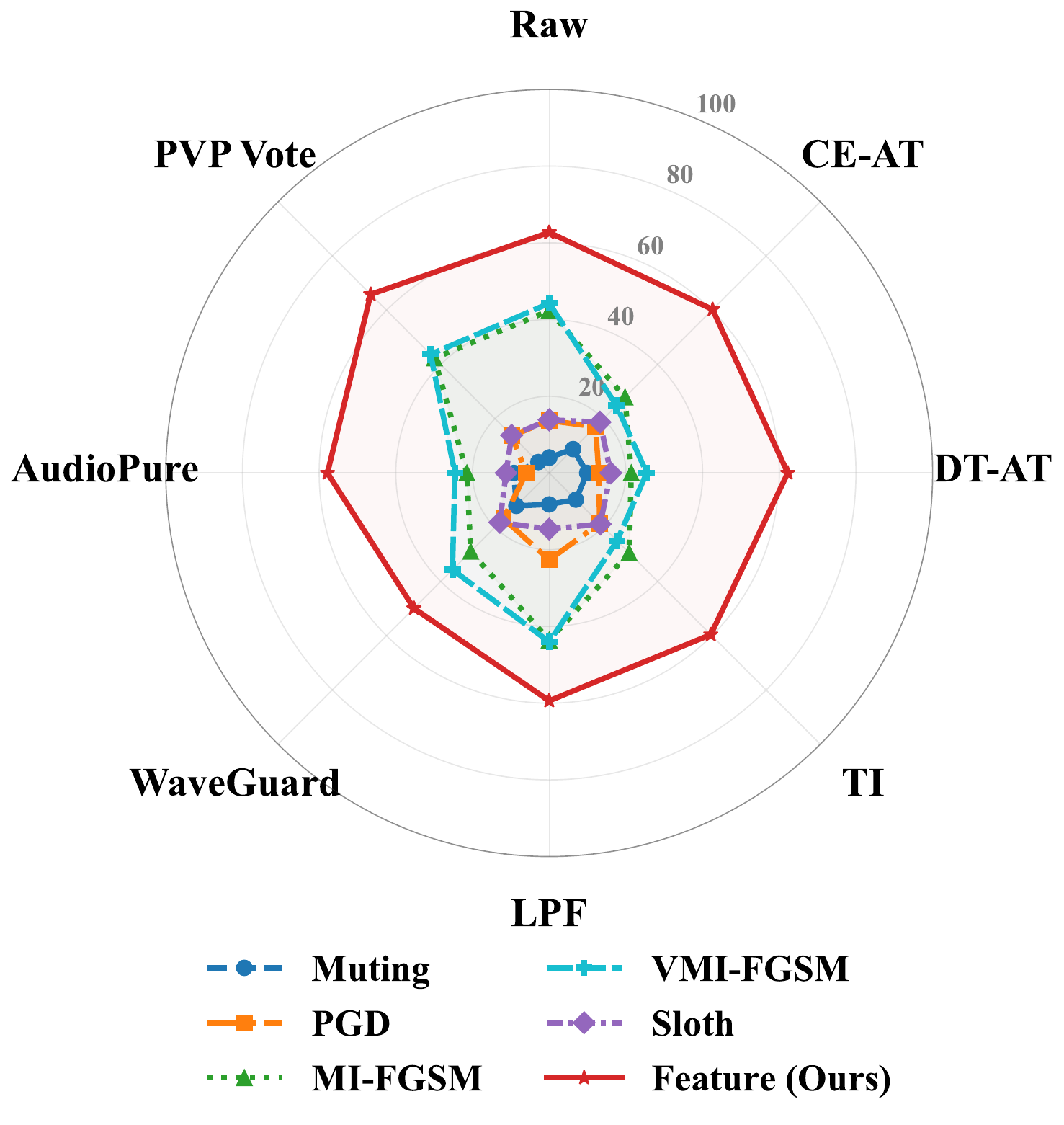}
        \caption{Whisper-base}
        \label{fig:whisper_transfer_base}
    \end{subfigure}

    \vspace{1mm}

    \begin{subfigure}[t]{0.49\linewidth}
        \centering
        \includegraphics[width=\linewidth]{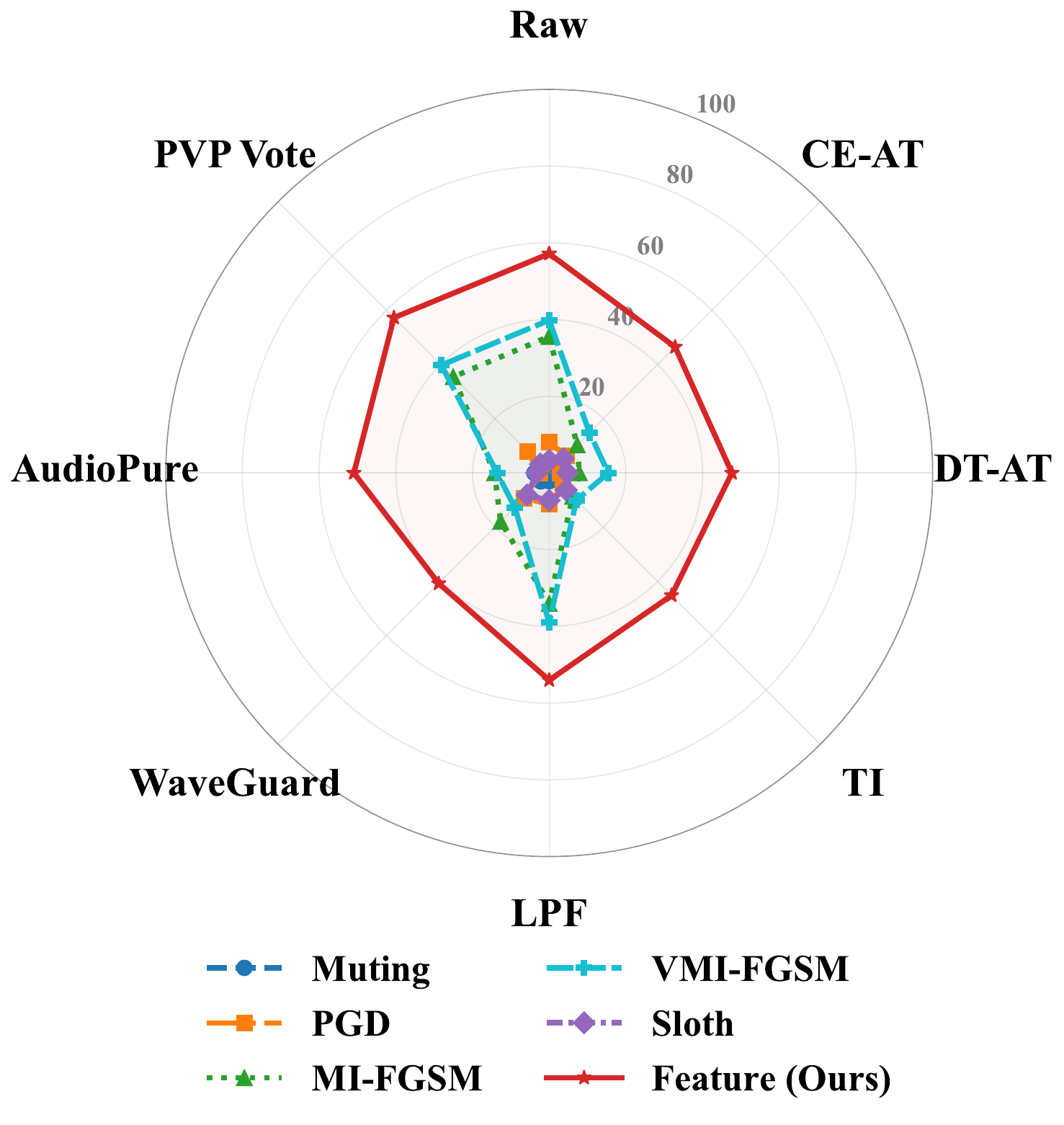}
        \caption{Whisper-medium}
        \label{fig:whisper_transfer_medium}
    \end{subfigure}
    \hfill
    \begin{subfigure}[t]{0.49\linewidth}
        \centering
        \includegraphics[width=\linewidth]{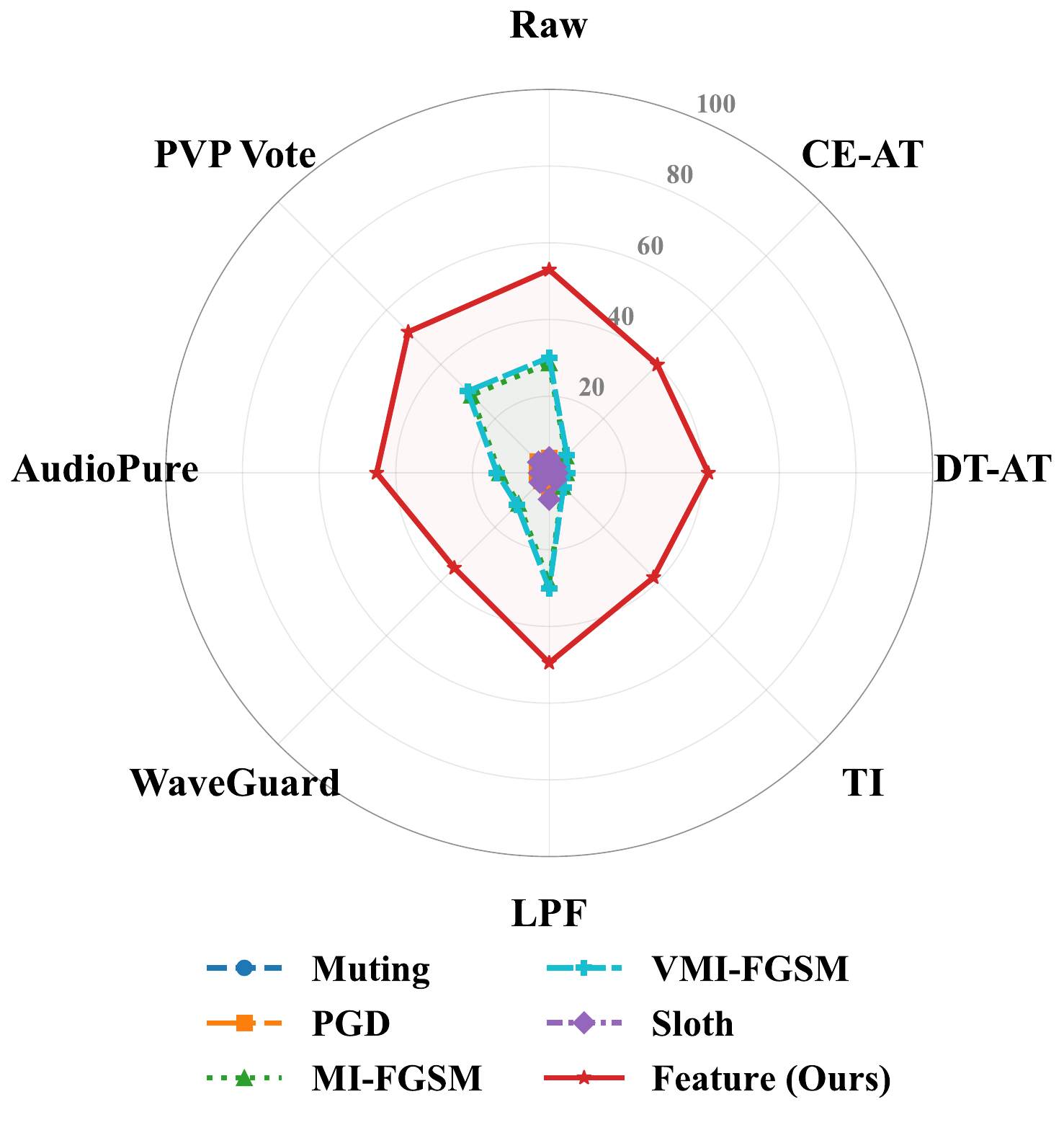}
        \caption{Whisper-large}
        \label{fig:whisper_transfer_large}
    \end{subfigure}
    
    \caption{\small Transferability across the Whisper model family.}
    \label{fig:whisper_family_transfer}
\end{figure}

\mypara{Transferability Evaluation Results}
We first evaluate whether adversarial examples generated on Whisper-small can transfer to other models within the Whisper family. 
Specifically, we use Whisper-small as the public surrogate to generate adversarial audio, and then evaluate the same audio on Whisper-tiny and Whisper-large under both adversarial training and input preprocessing defenses. As shown in ~\Cref{fig:whisper_family_transfer}, our Feature Attack consistently achieves the strongest transfer performance across both smaller and larger Whisper models. 
Existing waveform-level attacks, such as Muting, PGD, MI-FGSM, and Sloth, either perform moderately on the raw model or degrade substantially under defenses. 
In contrast, our attack remains effective across almost all defended settings. For instance, on Whisper-tiny, our method achieves 64.90\% WER under DT-AT and 54.75\% WER under WaveGuard, while the strongest baselines are much lower. On Whisper-large, which is substantially different in model capacity from the surrogate Whisper-small, our attack still achieves 52.96\% WER on the raw model and remains effective under defenses such as CE-AT, AudioPure, and PVP Vote. 
This indicates that feature-vocoder adversarial examples are not merely overfitted to the waveform-level gradients of Whisper-small, but can transfer across Whisper models with different capacities.

\begin{figure}[t]
    \centering
    \begin{subfigure}[t]{0.48\linewidth}
        \centering
        \includegraphics[width=\linewidth]{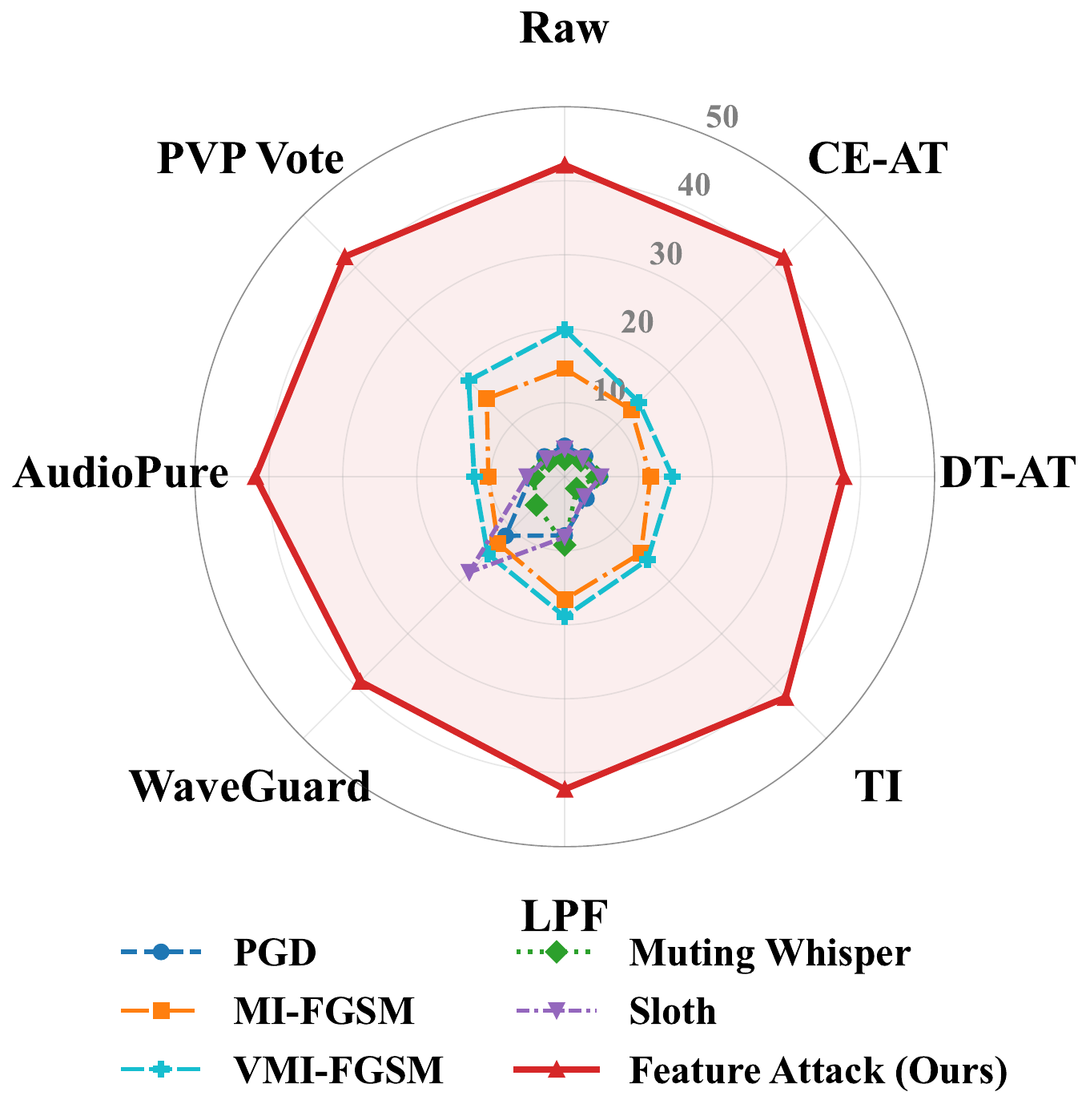}
        \caption{Wav2Vec2 CTC}
        \label{fig:ctc_transfer_wav2vec2}
    \end{subfigure}
    \hfill
    \begin{subfigure}[t]{0.48\linewidth}
        \centering
        \includegraphics[width=\linewidth]{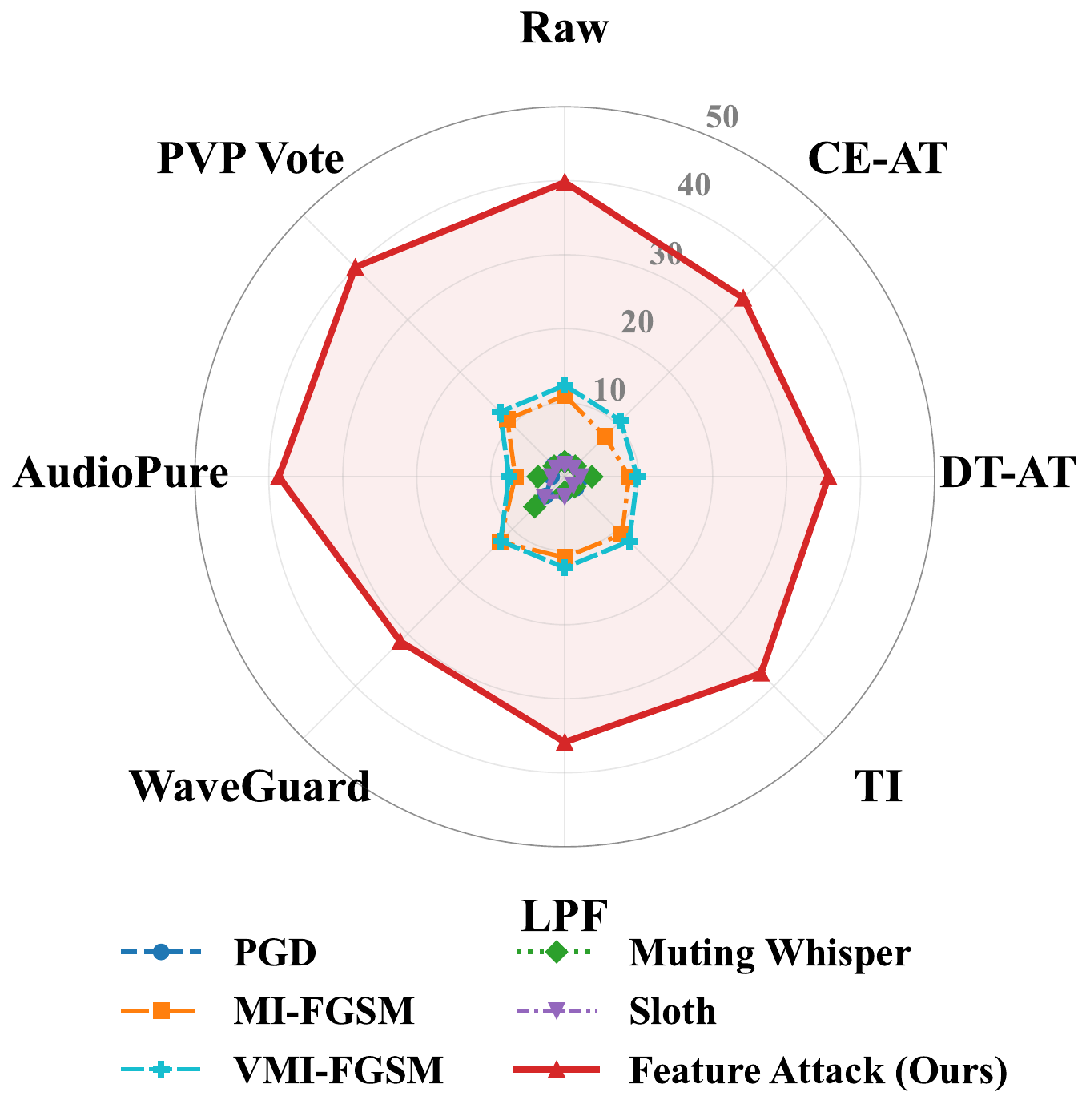}
        \caption{HuBERT CTC}
        \label{fig:ctc_transfer_hubert}
    \end{subfigure}
    
    \caption{\small Cross-architecture transferability to self-supervised CTC-based ASR models.}
    \label{fig:ctc_cross_architecture_transfer}
\end{figure}

We further evaluate cross-architecture transfer by applying adversarial examples generated on Whisper-small to self-supervised CTC-based ASR models, including HuBERT CTC and Wav2Vec2 CTC. This setting is more challenging because the target models use different acoustic encoders and decoding objectives from Whisper.
As shown in~\Cref{fig:ctc_cross_architecture_transfer}, most baseline attacks transfer poorly to these CTC models, with WER often remaining close to the clean or weakly attacked performance, especially under adversarial training and preprocessing defenses.
In contrast, our Feature Attack achieves consistently higher WER across both architectures and all defense settings.
For instance, on HuBERT CTC, our method reaches 39.80\% WER on the raw model and remains above 30\% WER under all defenses. On Wav2Vec2 CTC, our attack achieves 42.12\% WER on the raw model and 48.24\% WER under LPF, outperforming all waveform-level baselines by a large margin. This indicates that perturbing SSL feature trajectories and reconstructing them through a neural vocoder produces adversarial speech that transfers beyond the Whisper family, revealing vulnerabilities shared across different ASR architectures.

\begin{figure}[t]
    \centering
    \includegraphics[width=0.95\columnwidth]{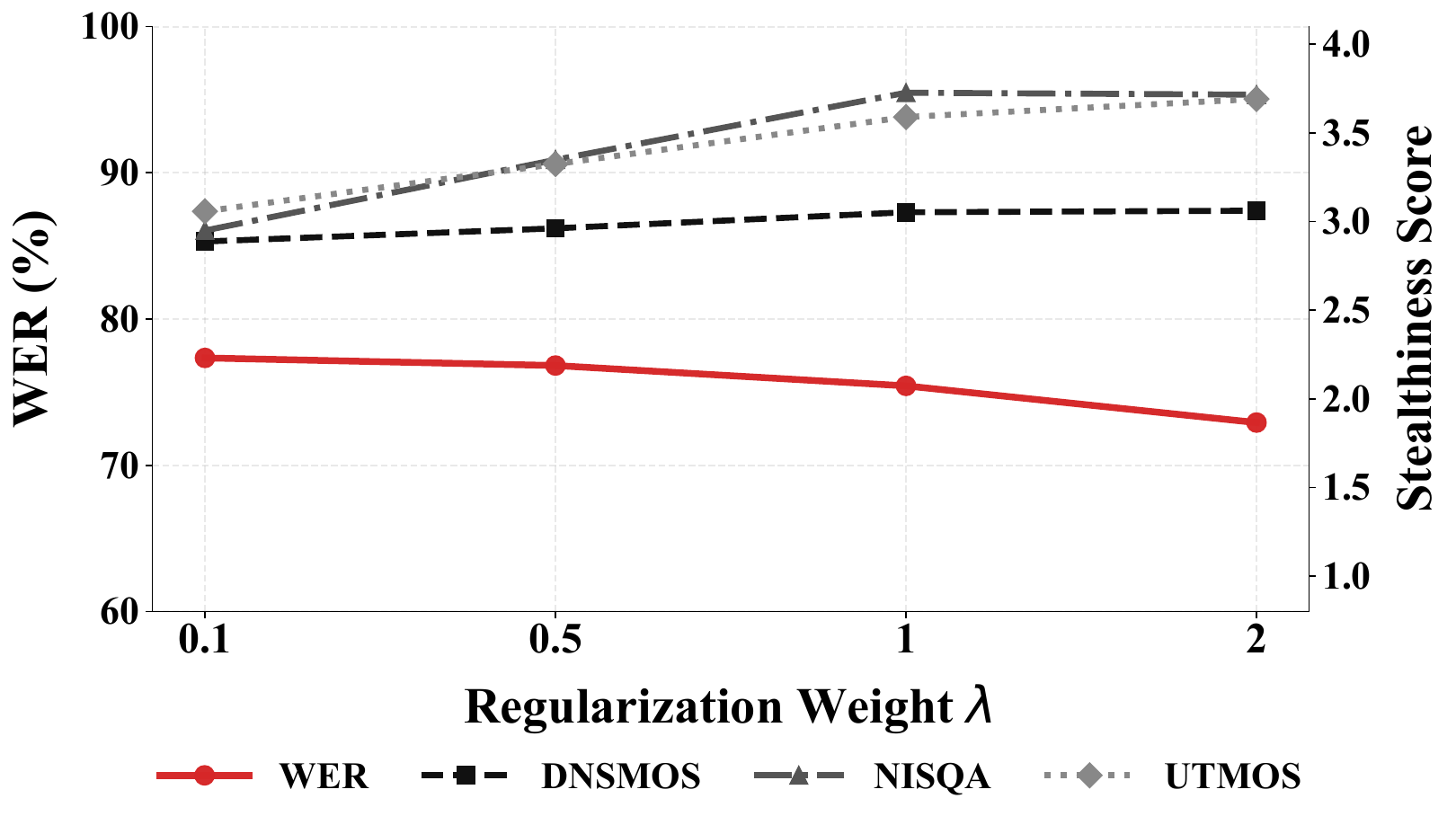}
    \caption{\small Ablation study on the $\lambda$ Selection.}
    \label{fig:lamda}
\end{figure}

\begin{figure}[t]
    \centering
    \includegraphics[width=0.95\columnwidth]{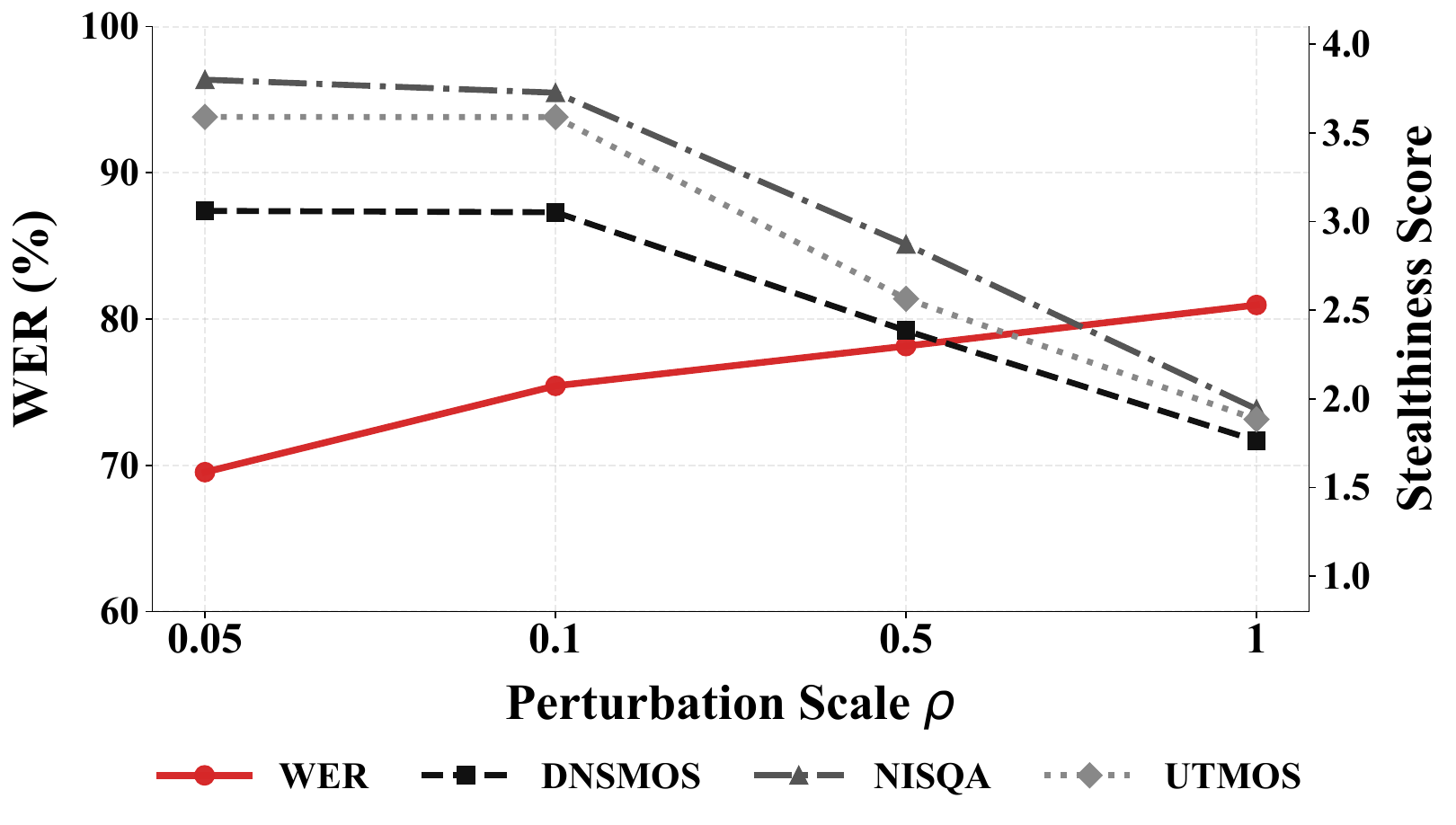}
    \caption{\small Ablation study on the $\rho$ Selection.}
    \label{fig:u}
\end{figure}

\mypara{Imperceptibility Results}
We evaluate the imperceptibility of our adversarial audio by sweeping two key hyperparameters in our objective: the perceptual regularization weight $\lambda_{\mathrm{perc}}$ and the normalized feature perturbation budget $\rho$. For each setting, we report both attack effectiveness, measured by WER, and perceptual speech quality, measured by DNSMOS~\cite{reddy2021dnsmos}, NISQA~\cite{mittag2021nisqa}, and UTMOS~\cite{saeki2022utmos}. The clean reference audio obtains 3.07 DNSMOS, 3.71 NISQA, and 3.76 UTMOS. 
Based on this trade-off, we choose $\lambda_{\mathrm{perc}}=1$ and $\rho=0.1$ as our default setting. Under this configuration, our adversarial audio achieves 75.43\% WER while maintaining perceptual quality close to the clean reference, with 2.95 DNSMOS, 3.53 NISQA, and 3.59 UTMOS. This indicates that the attack can induce substantial transcription errors without noticeably degrading speech quality.
The hyperparameter trends further justify this choice. As shown in ~\Cref{fig:lamda}, increasing $\lambda_{\mathrm{perc}}$ generally improves perceptual quality but gradually reduces attack strength, while a smaller $\lambda_{\mathrm{perc}}$ gives only marginally higher WER at the cost of weaker regularization. 
Similarly, as shown in ~\Cref{fig:u}, increasing $\rho$ can further increase WER, but it substantially degrades speech quality, especially under larger feature perturbation budgets. 
In contrast, $\rho=0.1$ provides a balanced operating point: it significantly improves attack effectiveness over a very small perturbation budget while keeping all perceptual metrics close to the clean audio. These results suggest that our selected hyperparameters achieve a favorable trade-off between adversarial effectiveness and imperceptibility.

We further conduct a human study to validate the perceptual stealthiness of our adversarial audio. 
We recruit ten volunteers and ask each participant to evaluate two independent sets of audio samples, with 100 samples in each set. 
In the first set, participants are presented with paired clean and adversarial audio clips and asked whether they can perceive any difference between them. 
In the second assessment, participants listen to adversarial audio clips alone, without access to the clean reference, and are asked to flag samples with noticeable distortion or suspicious artifacts. 
The results show that 86\% of the paired samples are judged to be indistinguishable between clean and adversarial audio, and 92\% of the adversarial samples are not perceived as having significant distortion in the single assessment. 
We also ask participants to manually transcribe the adversarial audio. The resulting human transcription WER is 5.47\%, indicating that the adversarial audio remains intelligible to human listeners despite causing substantial ASR errors.
These results are consistent with the objective perceptual metrics and indicate that our feature-vocoder attack preserves high perceptual quality while inducing substantial transcription errors.

\begin{figure}[t]
    \centering
    \includegraphics[width=0.8\columnwidth]{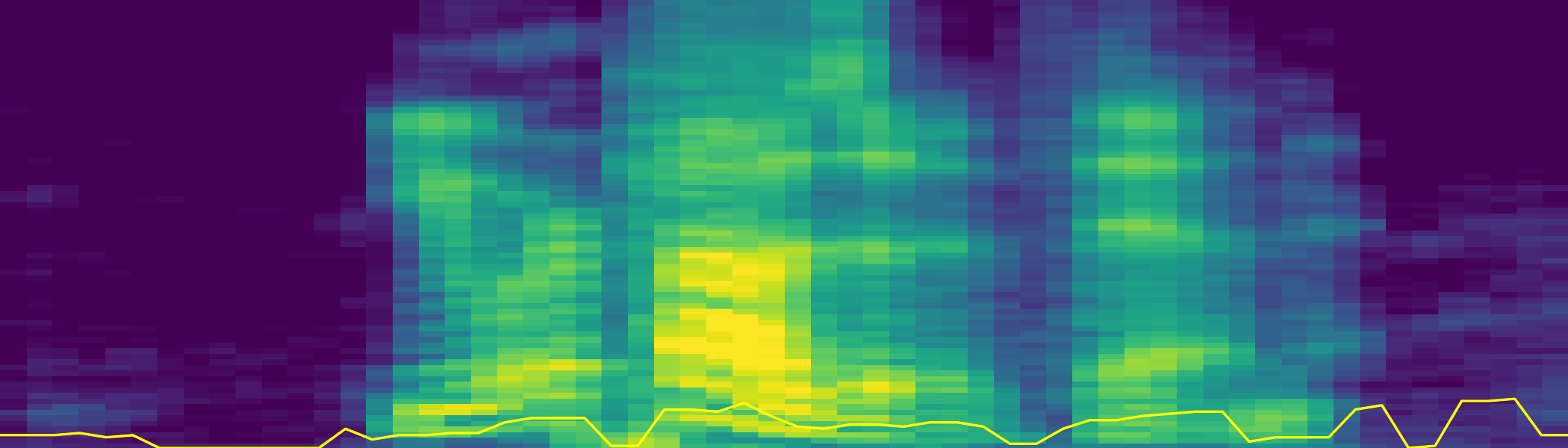}
    \vspace{1mm}
    \includegraphics[width=0.8\columnwidth]{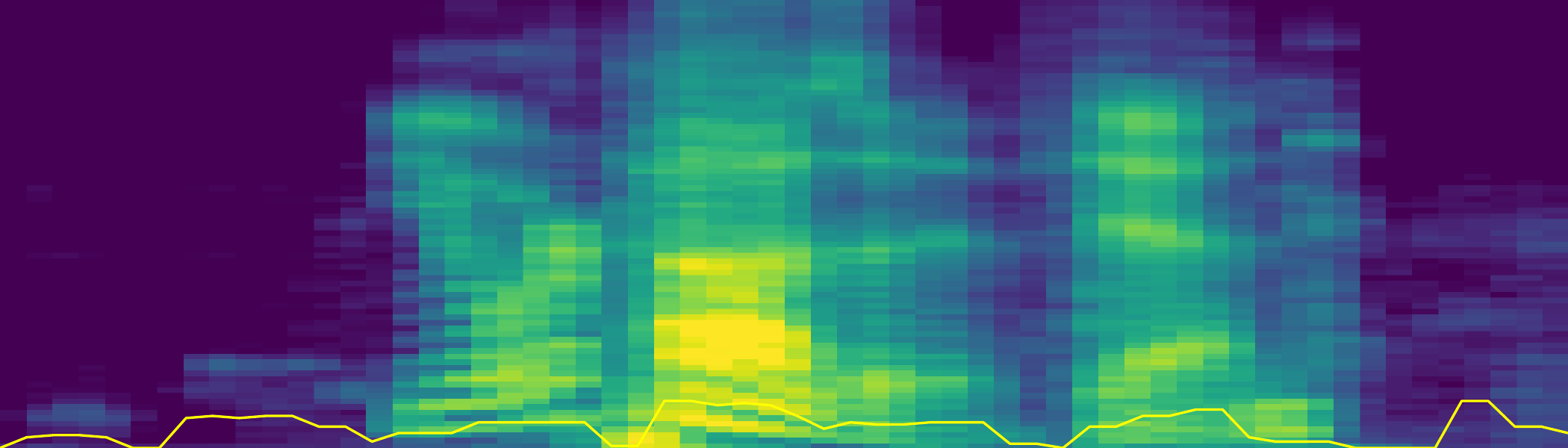}
    \caption{\small Mel-spectrogram and pitch contour comparison between the clean reference audio (above) and the corresponding adversarial audio (down). }
    \label{fig:mel_pitch_vis}
\end{figure}

To further analyze the imperceptibility of our adversarial audio, we visualize the Mel-spectrogram and pitch contour of a representative clean audio and its corresponding adversarial example. 
As shown in ~\Cref{fig:mel_pitch_vis}, the adversarial audio preserves the global time-frequency structure of the clean audio, including the main energy regions and pitch trajectory. Quantitatively, this sample achieves a high Mel-spectrogram correlation of 0.9733 between the clean and adversarial audio, indicating that the feature-vocoder attack largely maintains the acoustic structure of the original speech. 

\mypara{Loss Ablation Results}
We conduct a loss ablation study to understand the contribution of each component in our clean-referenced perceptual objective. Starting from the full objective, we remove one component at a time, including the ASR attack loss, the overall perceptual constraints, the temporal jitter regularization, and the frequency-energy regularization. 
We then evaluate both attack effectiveness, measured by WER, and perceptual quality, measured by DNSMOS. Without the ASR attack loss, the generated audio only reaches 6.6\% WER, which is close to the clean baseline and confirms that the attack objective is necessary for inducing transcription errors. The results show that the full objective achieves the best balance between attack strength and audio quality. 
Removing the perceptual constraints leads to both lower WER and degraded perceptual quality, suggesting that the perceptual loss is not merely a restrictive constraint but also helps guide the feature perturbation toward vocoder-compatible and speech-like directions. 
Removing either temporal jitter or frequency-energy regularization also weakens the attack and reduces speech quality, indicating that both terms are useful for stabilizing the reconstructed audio and avoiding abnormal artifacts. 
Overall, the ablation confirms that our loss design improves imperceptibility while also supporting stronger effective attacks after neural vocoding.
\begin{figure}[t]
    \centering
    \includegraphics[width=0.95\columnwidth]{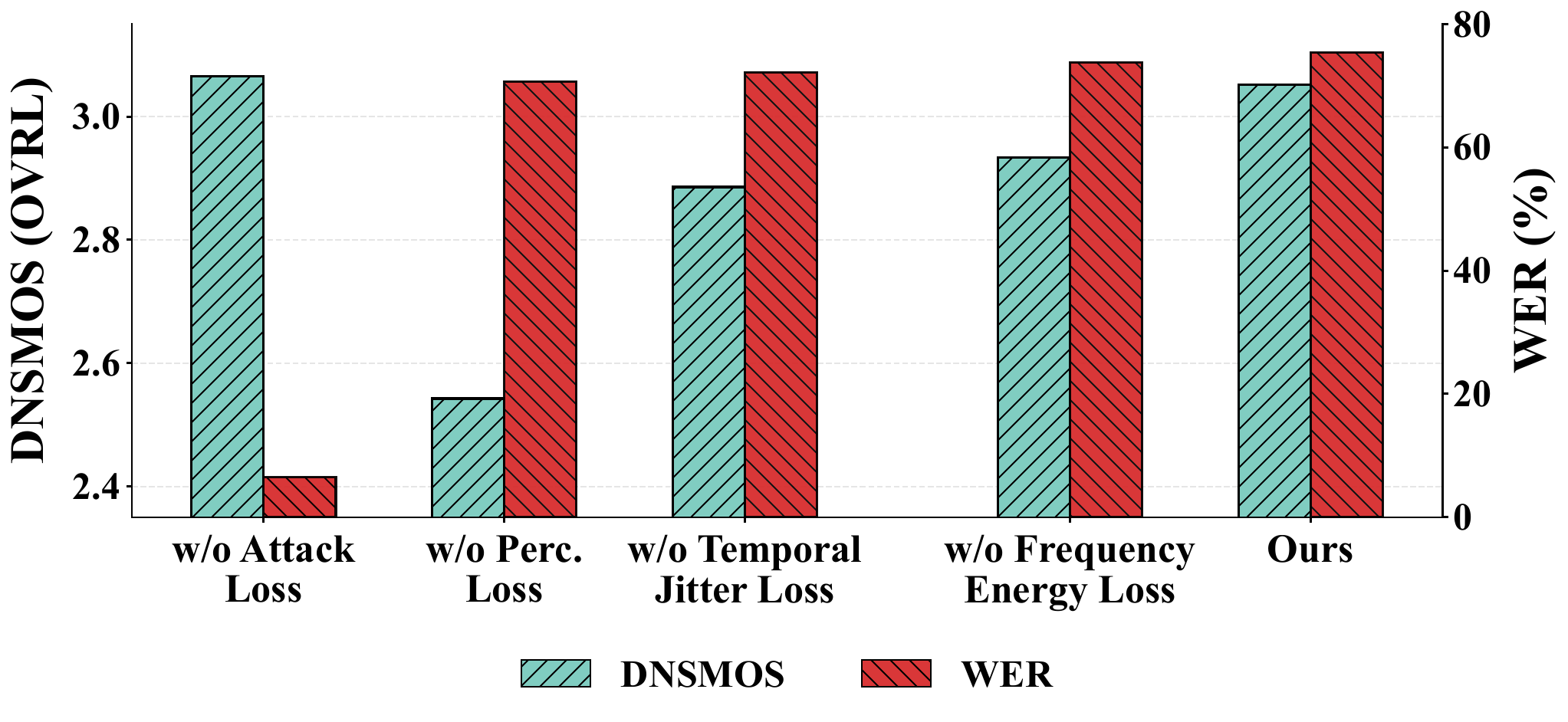}
    \caption{\small Ablation study on the loss.}
    \label{fig:loss}
\end{figure}

\mypara{Physical Experiment}
To examine whether our attack remains effective beyond the digital setting, we conduct a preliminary over-the-air (OTA) experiment with 100 English utterances
from three human speakers. The adversarial audio is generated using the same feature-vocoder attack pipeline, played through a loudspeaker, re-recorded by
smartphones, and then transcribed by the target ASR system. As shown in~\Cref{tab:physical_ota}, the clean OTA recordings obtain a WER of 7.45\%, indicating that the playback and re-recording pipeline itself does not severely degrade ASR performance. In contrast, the adversarial OTA recordings still achieve a WER of \textbf{78.23\%}. 
This large gap suggests that the proposed attack can retain substantial effectiveness under physical playback and re-recording conditions. 
We provide additional setup details in Appendix~\ref{sec:appendix-physical}.

\section{Conclusion}
In this paper, we propose a Clean-Referenced Feature-Vocoder Attack that moves adversarial optimization from raw waveforms to SSL speech representations and reconstructs the perturbed features into audio through a frozen vocoder. 
This design improves black-box transferability and avoids representing adversarial variation as explicit additive waveform noise.


\section{Limitations}

Our work has several limitations. Firstly, although our attack transfers across Whisper-family models and CTC-based architectures, the evaluation is still limited to a finite set of ASR systems and defenses. 
Future work should study whether similar feature-vocoder attacks generalize to broader commercial ASR services, streaming ASR systems, and larger multilingual speech foundation models. 
Secondly, our current implementation relies on a specific SSL encoder and neural vocoder. Different choices of SSL features, vocoder architectures, or reconstruction quality may affect both attack effectiveness and perceptual quality. Third, while our objective and human study suggest that the generated adversarial audio is perceptually similar to clean speech, imperceptibility remains difficult to fully characterize with automatic metrics alone. 
Larger-scale human studies under more diverse listening environments would provide a more comprehensive evaluation. Finally, our physical-world experiment demonstrates initial feasibility, but it is conducted with a limited number of speakers, devices, and acoustic conditions. More extensive over-the-air evaluations are needed to understand robustness under real-world playback, room reverberation, background noise, and device-specific recording pipelines.

\section{Ethical Considerations}

This work studies adversarial attacks on ASR systems to better understand and improve their robustness. 
The proposed attack could potentially be misused to degrade transcription systems or bypass defenses if applied irresponsibly. To mitigate this risk, we frame the study as a robustness evaluation tool and focus on exposing limitations of current waveform-bounded defenses rather than enabling real-world abuse. 
Our experiments are conducted on research datasets and controlled evaluation settings. The physical-world study is performed with volunteer participants, and no sensitive or private speech content is used. We encourage future use of this work for defensive purposes, such as evaluating ASR systems under broader threat models, developing feature-space-aware defenses, and improving adversarial training beyond waveform-level perturbations. 
Any deployment of the proposed method should follow responsible disclosure practices and avoid targeting real users or production ASR services without authorization.

\bibliographystyle{plain}
\bibliography{custom}

\appendix



\section{Baseline Implementation Details}
\label{sec:appendix-implementation}

Because the compared attacks operate in different spaces, exact norm matching is not meaningful across additive waveform perturbations, universal acoustic prefixes, and feature-vocoder transformations. We therefore use a common validation-based selection protocol under fixed perceptual-quality constraints.
For each attack, we sweep its attack-strength hyperparameters on a held-out validation split and discard configurations whose DNSMOS, NISQA, or UTMOS score is lower than 2.5. 
Among the remaining configurations, we select the one with the highest surrogate WER/CER on the validation split. 
Target model and defense results are not used during this selection.

\begin{equation}
    \theta_a^\star
    =
    \arg\max_{\theta \in \Theta_a}
    \operatorname{Err}
    \left(
    f_s(x_{\mathrm{adv}}^{a,\theta}), y
    \right)
\end{equation}
The selected configuration $\theta_a^\star$ is then fixed and used for all black-box target models and defense evaluations. Target model and defense results are never used for hyperparameter selection.

For iterative attacks, including PGD, MI-FGSM, VMI-FGSM, SlothSpeech, and our Feature Attack, we use the same optimization budget of 50 update steps. For our method, we use $\rho=0.1$, $\lambda_{\mathrm{perc}}=1$, and 50 optimization steps, selected by the same validation protocol. The complete hyperparameter search ranges, perceptual thresholds, and selected configurations are reported in~\Cref{tab:baseline_tuning_protocol,tab:perceptual_evaluation}.

\begin{table*}[t]
\centering
\small
\resizebox{\textwidth}{!}{
\begin{tabular}{l|l|l|l|l}
\toprule
\textbf{Attack}
& \textbf{Tuned Hyperparameters}
& \textbf{Validation Search Grid}
& \textbf{Selected Configuration}
& \textbf{Selection Rule} \\
\midrule

PGD
& Perturbation budget $\epsilon$, step size $\alpha$
& $\epsilon \in \{0.001, 0.002, 0.005, 0.01, 0.02\}$,
$\alpha \in \{\epsilon/50, 2\epsilon/50, \epsilon/10\}$
& $\epsilon=0.02$, $\alpha=2\epsilon/50$
& Max validation WER/CER on $f_s$ with DNSMOS, NISQA, UTMOS $>2.5$ \\

MI-FGSM
& Perturbation budget $\epsilon$, step size $\alpha$, momentum decay $\mu$
& $\epsilon \in \{0.001, 0.002, 0.005, 0.01, 0.02\}$,
$\alpha \in \{\epsilon/50, 2\epsilon/50, \epsilon/10\}$,
$\mu \in \{0.5, 1.0\}$
& $\epsilon=0.02$, $\alpha=2\epsilon/50$, $\mu=1.0$
& Max validation WER/CER on $f_s$ with DNSMOS, NISQA, UTMOS $>2.5$ \\

VMI-FGSM
& Perturbation budget $\epsilon$, step size $\alpha$, momentum decay $\mu$,
variance samples $N$, neighborhood scale $\beta$
& $\epsilon \in \{0.001, 0.002, 0.005, 0.01, 0.02\}$,
$\alpha \in \{\epsilon/50, 2\epsilon/50, \epsilon/10\}$,
$\mu \in \{1.0\}$,
$N \in \{3,5\}$,
$\beta \in \{1.0,1.5,2.0\}$
& $\epsilon=0.02$, $\alpha=2\epsilon/50$, $\mu=1.0$,
$N=5$, $\beta=1.5$
& Max validation WER/CER on $f_s$ with DNSMOS, NISQA, UTMOS $>2.5$ \\

Muting Whisper
& Prefix length, prefix amplitude scale
& Prefix length $\in \{5120, 10240, 20480\}$ samples
$(0.32\mathrm{s}, 0.64\mathrm{s}, 1.28\mathrm{s}$ at 16 kHz),
amplitude scale $\in \{0.005, 0.01, 0.02\}$
& Prefix length $=0.64s$, amplitude scale $=0.02$
& Max validation WER/CER on $f_s$ with DNSMOS, NISQA, UTMOS $>2.5$ \\

SlothSpeech
& Distance criterion, learning rate, distance regularization weight
& Distance criterion $\in \{\ell_2,\ell_{\infty}\}$,
learning rate $\in \{10^{-2}, 5{\times}10^{-2}, 10^{-1}\}$,
distance factor $\in \{0.05, 0.1, 0.2\}$
& Distance criterion $=\ell_2$, learning rate $=10^{-1}$,
distance factor $=0.1$
& Max validation WER/CER on $f_s$ with DNSMOS, NISQA, UTMOS $>2.5$ \\

Feature Attack (Ours)
& Feature budget $\rho$, perceptual weight $\lambda_{\mathrm{perc}}$
& $\rho \in \{0.05, 0.1, 0.5, 1.0\}$,
$\lambda_{\mathrm{perc}} \in \{0.1, 0.5, 1.0, 2.0\}$
& $\rho=0.1$, $\lambda_{\mathrm{perc}}=1.0$
& Max validation WER/CER on $f_s$ with DNSMOS, NISQA, UTMOS $>2.5$ \\

\bottomrule
\end{tabular}
}
\caption{
Hyperparameter tuning protocol for all attacks. All hyperparameters are selected on a held-out validation split using only the surrogate model $f_s$ and objective speech-quality metrics. 
Configurations whose DNSMOS, NISQA, or UTMOS score is below 2.5 are discarded before selecting the strongest validation attack. Target models and defenses are never used during hyperparameter tuning. The selected configuration for each attack is fixed before black-box evaluation.
}
\label{tab:baseline_tuning_protocol}
\end{table*}

\section{Physical-World Evaluation Details}
\label{sec:appendix-physical}

We conduct an over-the-air (OTA) evaluation to test whether the proposed feature-vocoder adversarial examples remain effective after physical playback and re-recording. 
The experiment uses 100 English utterances recorded from three human speakers. For each utterance, we generate adversarial audio using the same feature-vocoder attack pipeline as in the digital experiments. The generated audio is then played through a loudspeaker and re-recorded by smartphones before being transcribed by the target ASR system.

All recordings are conducted in an indoor environment. During recording, the phone microphone is placed facing the loudspeaker, and the playback volume is kept fixed across trials. This setup introduces common physical-channel distortions, including loudspeaker response, room acoustics, smartphone microphone response, and device-specific recording effects.

We evaluate two OTA conditions: clean OTA audio and adversarial OTA audio. 
The clean OTA condition serves as a control for ASR errors caused by playback and re-recording alone. As shown in~\Cref{tab:physical_ota}, clean OTA audio obtains a WER of 7.45\%, while adversarial OTA audio obtains a WER of \textbf{78.23\%}. The large increase in WER indicates that the adversarial effect is not explained by the physical recording pipeline alone. 
We view this result as an initial physical-world validation; more extensive evaluation across different rooms, devices, distances, playback volumes, and background-noise conditions is left for future work.
\begin{table}[t]
\centering
\small
\begin{tabular}{l|c}
\toprule
\textbf{Condition} & \textbf{WER (\%)} \\
\midrule
Clean OTA audio & 7.45 \\
Adversarial OTA audio & \textbf{78.23} \\
\bottomrule
\end{tabular}
\caption{
Physical-world over-the-air evaluation. Audio is played through a loudspeaker and re-recorded by smartphones before ASR transcription. Clean OTA audio serves as a control for errors introduced by the playback and re-recording pipeline.
}
\label{tab:physical_ota}
\end{table}

\begin{table}[t]
\centering
\resizebox{\columnwidth}{!}{
\begin{tabular}{l|c|ccc}
\toprule
\multirow{2}{*}{\textbf{Attack Method}} 
& \multirow{2}{*}{\textbf{Attack Domain}} 
& \multicolumn{3}{c}{\textbf{Perceptual Quality (MOS $\uparrow$)}} \\
\cmidrule(lr){3-5}
& & \textbf{DNSMOS} & \textbf{NISQA} & \textbf{UTMOS} \\
\midrule
Clean & --- & 3.075 & 3.714 & 3.755 \\
\midrule
PGD & Waveform & 2.761 & 3.094 & 3.314 \\
MI-FGSM & Waveform & 2.804 & 2.523 & 2.518 \\
VMI-FGSM & Waveform & 2.763 & 2.517 & 2.528 \\
Muting Whisper & Prefix & \textbf{3.011} & \textbf{3.640} & \textbf{3.729} \\
SlothSpeech & Waveform & 2.895 & 3.152 & 3.458 \\
\midrule
\rowcolor{gray!10}
Feature Attack (Ours) & Feature-vocoder & \underline{2.954} & \underline{3.526} & \underline{3.590} \\
\bottomrule
\end{tabular}
}
\caption{Objective acoustic perceptual quality evaluation of various adversarial audio attacks. The results demonstrate that our Feature Attack preserves high objective speech quality (MOS) compared to baseline waveform attacks that perturb the entire sequence.}
\label{tab:perceptual_evaluation}
\end{table}  

\section{Artifacts and Licenses.}
We use publicly available research artifacts, including LibriSpeech, AISHELL-1, Whisper, WavLM, HuBERT, Wav2Vec2, and HiFi-GAN, following their respective licenses and terms of use. 
All datasets are used only for research evaluation, and we do not redistribute the original datasets or model checkpoints.

\section{Computational Budget.}
All attacks are generated on a single NVIDIA L20 GPU (46GB). Our attack uses 50 optimization steps per utterance, with an average generation time of approximately 10 seconds per utterance. 
For LibriSpeech, train-clean-100 is used only for training-related preparation and fine-tuning where applicable, while the computational budget reported here is computed on the held-out 999-utterance subset of test-clean used for attack evaluation. 
Generating adversarial examples for this test-clean subset takes about 2.8 GPU-hours; including model loading, I/O overhead, and evaluation runs, we conservatively report approximately 4 GPU-hours.

\end{document}